\documentclass[preprint]{revtex4}
\newcommand{\beq}{\begin{equation}}

\newcommand{\eeq}{\end{equation}}
\newcommand{\barray}{\begin{eqnarray}}
\newcommand{\earray}{\end{eqnarray}}

\begin{document}
\title{A Class of Parameter Dependent Commuting Matrices}
\author{B Sriram Shastry }
\address{Department of Physics, University of California, Santa Cruz, CA 95064}
\date{\today}

\begin{abstract}
We present a  novel class of real symmetric  matrices in arbitrary dimension $d$,   linearly dependent on a parameter $x$. 
The matrix elements satisfy a set of nontrivial constraints that arise from asking for commutation of pairs of such matrices for all  $x$,  and an intuitive sufficiency condition for the solvability of certain linear equations that arise therefrom. This class of matrices generically violate the Wigner von Neumann non crossing rule, and  is argued to be intimately connected with  finite dimensional Hamiltonians of quantum integrable systems. 
\end{abstract}
\maketitle
\vspace{.5in}

{\bf Introduction} We present a novel  class of real symmetric  matrices in arbitrary finite dimensions $d$. These  matrices  are   linearly dependent on a parameter $x$,  which  plays the role of an interaction constant for a quantum system.  The matrix elements satisfy a set of  linear as well as non linear constraints that are derived. Each matrix $\alpha(x)$ ( as in 
Eq(\ref{alphax})) of this class has  commuting partner matrices $\beta(x)$  ( as in Eq(\ref{betax})), also linearly dependent on $x$, and having  $d+1$ independently assignable real parameters. This class of matrices generically exhibits   level crossings as $x$ is varied, and is intimately connected with  finite dimensional Hamiltonians of integrable systems with {\em dynamical symmetries}.  

The study of linearly parameter dependent matrices is very popular in the context of quantum chaos \cite{berry87,berry83,bohigas,guhr}, where it  models the change in universalty class of quantum systems, or the temporal variation of correlations within the same class via the matrix Brownian motion model of Dyson\cite{dyson,altshuler,narayan}. This work has a different goal, that of formulating constraints so that the system remains `regular'' despite mixing with  another matrix. In fact such linearly parameter dependent matrices are prototypes of quantum integrable models. The well known case of the anisotropic Heisenberg model solved by  Bethe's Ansatz \cite{bethe} possesses ``infinite'' higher conservation laws, {\em every one of them being linear in the anisotropy parameter}\cite{thacker}.

The immediate motivation for our work comes from a study of the parameter dependence of eigenvalues of blocks of finite dimensional matrices arising  from completely integrable  models of interacting  quantum systems, such as the Heisenberg model\cite{bethe} or the  1-d Hubbard model\cite{hubbard}. The famous Wigner- von Neumann (WvN)\cite{vwn} non crossing rule for parametric evolution  of eigenvalues in quantum mechanics is a fundamental guiding principle for understanding level crossings, with violations or exceptions being termed as ``accidental degeneracies''. The term accident is used since usually there is no specific  ``space time symmetry''  responsible for such a degeneracy.  There is a general belief  that  {\em dynamical symmetries}, i.e. operators {\em dependent on the interaction parameter} also lead to degeneracies and  violations of the WvN rules, despite the lack   of a general group theoretic argument of the type that space time symmetries allow\cite{lenz}.

Dynamical symmetries occur in most integrable systems.  Our interest is also in sharpening the notion of complete or exact  integrability in the context of finite dimensional systems. In classical mechanics we have a very clear statement about  the meaning of complete integrability, namely that the number of degrees of freedom equals the number of  conservation laws that are mutually consistent. The meaning of the term ``degrees of freedom'' is quite unambiguous. For example   in the trivial case of a set of harmonically coupled oscillators, it is the number of oscillators so coupled. In quantum mechanics, we do have similar unambiguous models, such as the Calogero-Sutherland-Moser system\cite{calogero} in the continuum, or the Toda\cite{toda} lattice, where there  exists a natural definition of a degree of freedom, in complete parallel to the classical situation,  essentially the number of ``particle'' type variables. However, in the case of other quantum integrable models that arise in condensed matter physics, such as the Heisenberg spin chain or the 1-d Hubbard model,
  the situation is ambiguous: the number of spin flips, or particles is   variable.  These models are defined on a discrete lattice, and have a state space that is {\em  in general finite}, and lead to finite dimensional matrices depending upon a parameter, say the spin space anisotropy or the interaction strength. Presented with a finite dimensional matrix arising by restricting  such a model to a finite lattice, it is challenging to distinguish it from other matrices of the same dimension. Without  referring  back to the defining parent models, it is generally impossible to recognize their being ``integrable'', whatever that word implies! Indeed an extreme and  skeptical view would  challenge the very notion  of complete integrability in  a finite dimensional setting. One may argue that there are always $d$ commuting independent matrices for {\em any given matrix} $H$; one merely diagonalizes the matrix and in its eigenbasis, constructs the $d$ projection operators $P_j= |j><j|$, so that $[H,P_j]=0=[P_i,P_j]$. 

Our  viewpoint in this work is that the {\em parameter dependence of eigenvalues contains the essence of  quantum integrability} for such finite dimensional models. These lead to a violation of the WvN rules, and hence to Poisson statistics of the energy level separation. While the relationship between dynamical symmetries and levels crossings is not yet precisely established, there are several studies that indicate a deep relationship between them. For example,
 the beautiful numerical work of Heilmann and Lieb in 1971\cite{heilmann_lieb} on the 1-dimensional Hubbard model on a six site ring, shows that after { all the known space time and internal space symmetries} are carefully extracted, the finite dimensional blocks of matrices labelled by the appropriate quantum numbers, displaying scant regard for the WvN rules,   have a  large number of level crossings as the  interaction constant  is varied. More recent work of Yuzbashyan {\it et al}\cite{emil} has examined the detailed connection between these level crossings and the dynamical symmetry operators  of the Hubbard model found in 1986 by Shastry\cite{shas}, and provided considerable insight into this phenomenon. In particular, there is an explicit algebraic  demonstration for $d=3$ that dynamical symmetries definitely imply level crossings.

 Another related and  prominent manifestation of integrability is that  the energy level statistics of such  models display  Poisson type behavior  that allows  spacings to be arbitrarily small\cite{poilblanc}. This is in sharp contrast with the level spacings of generic (i.e. nonintegrable) models  
that exhibit level repulsion as expected from the WvN rule, and follow one of the three typical behaviors relevant to their class of quantum systems- namely the Gaussian orthogonal, unitary or symplectic classes\cite{mehta}.

It remains however, to state explicit  conditions on matrices in  arbitrary finite dimensions, that could  identify the proclivity for level crossings, and hence presumably define  completely integrable cases without reference to a  parent quantum model. This  goal is achieved  in this work,  we present    a set of conditions, and a class of matrices satisfying them in any dimension $d$. We find through examples that this class of matrices automatically leads to Poisson type statistics for the energy level separation, and also an abundance of level crossings.

 Our main results follow from asking for the conditions for two  parameter dependent matrices to commute with each other for all values of the parameter. Upon imposing a very intuitive sufficiency condition of autonomy (Type I matrices as explained  below), it leads to a set of constraints for each of the matrices. The partner matrix in the commutating pair is automatically also a member of the same class.

{\bf The matrix equations:} We consider real symmetric  matrices in d dimensions, linearly dependent on a real  parameter $x$ through 
\beq
\alpha(x) = a + x A \label{alphax}
\eeq
where $a$ is  a diagonal matrix with diagonal entries $\{a_1, a_2,....,a_d \}$ and $A$ is a real symmetric matrix\cite{identitymatrix}.
In the matrix $\alpha$ we have $d(d-1)/2$ off diagonal variables $A_{i,j}$, $d$ variables $a_j$ and a further $d$ variables $A_j \equiv A_{j,j}$, in addition to the real variable $x$. 
In an identical fashion we consider another matrix
\beq
\beta(x) = b + x B, \label{betax}
\eeq
where the diagonal matrix $b$ has entries $\{b_1,b_2,..,b_d \}$ and $B$ is another real symmetric matrix. Clearly at $x=0$ the matrices $\alpha$ and $\beta$ commute. We now ask the question, what are the conditions under which these commute for
{\em arbitrary values of $x$}? This clearly leads to two independent conditions
\beq
[a,B]  =  [b,A]  \;\mbox{ and} \;\;
\left[A,B\right]  =  0. \label{comm2} 
\eeq

The first set of  $d(d-1)/2 $ conditions,   are  expressible in terms of an  
antisymmetric matrix $S_{i,j}$ as follows:
\beq
 S_{i,j}  = \frac{A_{i,j}}{a_i-a_j} 
   =  \frac{B_{i,j}}{b_i-b_j}    \;\; \mbox{for}     \; \; {i < j.} \label{offdiagonals}
\eeq
The second set of equations can be written compactly using the Eq(\ref{offdiagonals}) in terms of
\beq
Y_{i,j}[\alpha] \equiv p_{i,j} - \frac{1}{ S_{i,j} ( a_i-a_j)} \sum_{l \neq i,j} S_{i,l} S_{l,j} (a_l-a_j) \label{yeq}
\eeq
where 
$
p_{i,j}  = \frac{A_i - A_j}{a_i - a_j}.
$
The $d(d-1)/2$ equations(\ref{comm2}) can be written as
\beq
Y_{i,j}[\alpha] =Y_{i,j}[\beta], \label{yeqy}
\eeq
where $Y_{i,j}[\beta]$ is obtained from the same formula Eq(\ref{yeq}), with $(a_j,A_j) \leftrightarrow (b_j,B_j)$.
Although it is possible to symmetrize $Y_{i,j}$  in $i \& j$, for future use it is better to use the present unsymmetric form. As expected the
Equation(\ref{yeqy}) is symmetric in $\alpha$ and $\beta$.  At this point, we disturb this symmetry, and  rearrange terms  so that we can solve for $\beta$  if $\alpha$ were given.
This implies with $ (b \wedge a)_{i,j,l} \equiv \left[ ( b_i a_j - b_j a_i) + ( b_j a_l - b_l a_j)+(b_l a_i - b_i a_l) \right]$
\barray
 \xi_{i,j}  \equiv   B_i-B_j 
 =  p_{i,j} ( b_i - b_j) + \frac{1}{a_i-a_j} \sum_{l \neq i,j} \frac{S_{i,l} S_{l,j} }{S_{i,j}} (b \wedge a)_{i,j,l}. \label{xidetail}  
\earray
This  set of  equations is linear in the  $ 2 d$ variables $\{b_1,b_2,..b_d \}$ 
and  $\{B_1, B_2, B_3,...,B_d\}$, and provides
 conditions for the matrix $\beta(x)$ to commute  with $\alpha(x)$ for all $x$. The offdiagonals of the matrix $ B$ namely $B_{i,j}$ 
are already fixed by Eq(\ref{offdiagonals}) once the $b_j's$ are picked in terms of the $S_{i,j}$.  For a given $\alpha(x) $ then, we have the freedom of choosing these $2 d$ variables subject to the $d(d-1)/2$ constraints Eq(\ref{xidetail}), which for large $d$,  are
many more than the number of variables available. These constraints are best expressed in terms of a ``triangle law'' for any three distinct indices $i,j,k$:
\beq
C(i,j,k) \equiv \xi_{i,j} + \xi_{j,k}+ \xi_{k,i}
 = 0. \label{constraints}
\eeq
Note that $C$ is antisymmetric under exchange of any pair of arguments. 
Since these constraints are linear in $b_j's$ we collect the coefficients and write 
\barray
C(i,j,k) =  \mu(i;j,k) b_i + \mu(j;k,i) b_j + \mu(k;i,j) b_k  + \sum_{l \neq i,j,k} \nu( l; i,j,k) b_l . \label{triples}
\earray
 $\mu$ is defined   below in Equation(\ref{mudef}) and
\beq
\nu(l;i,j,k)=\frac{S_{i,l} S_{l,j}}{S_{i,j}}+ \frac{S_{j,l} S_{l,k}}{S_{j,k}}+\frac{S_{k,l}S_{l,i}}{S_{k,i}}. \label{nueq}
\eeq

The solutions of Equations(\ref{constraints}) may be classified as being of two types. {\bf Type I} solutions correspond to
requiring the coefficients of every $b_r$  in Equation(\ref{triples}) to vanish individually. Such a choice is sufficient without being necessary, and gives us  constraints on the $S$ matrix all by itself. It also leads to {\em autonomous} constraints on $\alpha$, i.e. 
constraints involving the variables $a_j,A_j, S_{i,j}$, but  not  the $\beta$ variables. {\bf Type II} solutions  include all other 
possibilities, where the coefficients of $b_r$ are not all vanishing. These include the trivial solution $\alpha=\beta$, and are  less interesting as such. Type I solutions give us families of matrices $\alpha$ as we show below, that are definable in terms of 
relations involving $S$'s themselves,  and between these and the remaining variables.

Continuing our study of Type I solutions, 
we therefore equate the coefficients of $b's$ individually to zero, giving for all $^dC_3$ distinct triples $i,j,k$ the three index formulas, 
\barray
\mu(i;j,k)&=&0  \mbox{ plus  cyclic permutations of } i,j,k. \label{muall}
\earray
Likewise  we get  $^dC_4$ four index formulas      for each  distinct quadruple of indices,  $\nu(l;i,j,k) =0$, or rearranging a bit:
\beq
 S_{i,l} S_{l,j}S_{j,k}S_{k,i}+S_{j,l} S_{l,k}S_{k,i}S_{i,j}+S_{k,l}S_{l,i}S_{i,j}S_{j,k}  = 0. \label{ssss}
\eeq
Note that the  expression on the LHS is {\em fully symmetric} in the four variables. Although this relation involves quartics  in the $S$'s, inspection shows that for the case of real symmetric matrices, a considerable simplification occurs and 
 this constraint {  can be written in terms of bilinears}  if one inverts the { matrix elements} ( not the matrix itself!)  of $S$ and defines
\beq
R_{i,j}=\frac{1}{S_{i,j}}.
\eeq
  The vanishing of $\nu$ in Equation(\ref{ssss}) can be restated as the ( fermionic Wick's theorem type) requirement that the totally antisymmetric symbol $\phi$ vanishes for all distinct quadruples $i,j,k,l$ 
\barray
\phi_{(i,j,k,l)} &\equiv & R_{i,j} R_{k,l} - R_{i,k} R_{j,l} + R_{i,l}R_{j,k} = 0.  \label{rrrr}
\earray

For large $d$, the four index constraints Eq(\ref{rrrr}) are $\sim d^4/4!$ in number, representing a huge overdetermination since  the number of variables available, namely the $S_{i,j}$ are only $\sim d^2/2$ in number\cite{overdetermined}. Fortunately these identities are not all independent, and there exists an important extra identity relating {\em a set of five indices} that can be proved.  For any distinct  set of five indices $i,j,k,l,m$, we can easily see that
\beq
\phi_{(i,j,k,l)} R_{l,m} = - \phi_{(i,j,l,m)} R_{l,k} + \phi_{(i,k,l,m)} R_{l,j} - \phi_{(j,k,l,m)} R_{l,i}. \label{fiveindex}
\eeq
Using this {\em five index identity}  we can show that the number of independent quadruple relations are only $ (d-2)(d-3)/2$ in number. These may be chosen to be
\beq
\phi_{(1,2,3,4)}  =  0,\;
\phi_{(1,2,3,5)}  =  0,...\;
\phi_{(1,2,d-1,d)}  =  0. \label{setoffives} 
\eeq
From these equations, we can satisfy all others using Eq(\ref{fiveindex}) repeatedly. As an example consider $\phi_{(3,4,5,6)}$, we merely multiply by $R_{3,1}$ so that on using Eq(\ref{fiveindex}), we find
\beq
\phi_{(3,4,5,6)} R_{3,1}= \phi_{(3,1,5,6)} R_{3,4}  - \phi_{(3,1,4,6)} R_{3,5} + \phi_{(3,1,4,5)} R_{3,6},
\eeq
and each of these can be similarly   processed further, e.g.
\beq
\phi_{(1,3,4,6)} R_{1,2}= \phi_{( 1,2,4,6)} R_{1,3} - \phi_{(1,2,3,6)} R_{1,4} + \phi_{( 1,2,3,4)} R_{1,6}
\eeq
and reduced to forms involving $\phi_{(1,2,l,m)}$ which vanish according to  our list of Eq(\ref{setoffives}).

We now turn to the study of $\mu(i;j,k)$ which is written  compactly  as
\beq
\mu(i;j,k)  = Y_{i,j}- Y_{i,k} - \frac{S_{i,j} S_{i,k}}{S_{j,k}} \label{mudef} 
\eeq
 involving {\em the same variable} $Y_{i,j}$  that we encountered in Equation(\ref{yeq})\cite{alphadep}. The function $\mu$  is antisymmetric in $j,k$ as it stands. It can be antisymmetrized in all three variables by defining $\tilde{\mu}(i,j,k) \equiv  ( a_i -a_j) ( a_i -a_k) \mu(i;j,k)$, which satisfies
\barray
\tilde{\mu}(i,j,k)   
&& =(a_i - a_j)[ A_k- \frac{A_{i,k} A_{j,k}}{A_{i,j}}] 
+ (a_j-a_k) [ A_i- \frac{A_{i,j} A_{i,k}}{A_{k,j}}] \nonumber \\
&& +(a_k - a_i)[ A_j- \frac{A_{j,k} A_{k,i}}{A_{k,i}}]    - (a_i-a_j) ( a_i - a_k) \sum_{r \neq i,j,k} S_{i,r}\left( \frac{A_{j,r}}{A_{i,j}}- \frac{A_{k,r}}{A_{i,k}}\right). \label{muhat2}
\earray

We comment on some important properties of these formulae. The first part of Eq(\ref{muhat2}) consisting of three terms is explicitly antisymmetric in the three indices. The last line is antisymmetric in $j,k$ but not manifestly so  in $i,j$. We add to it 
a term with $i,j$ exchanged, leading to
\barray
\tilde{\mu}(i,j,k)+\tilde{\mu}(j,i,k)  = \sum_{r \neq i,j,k} (a_i-a_j) ( a_k- a_r) [ \frac{S_{i,r} S_{r,j}}{S_{i,j}}+ 
 \frac{S_{j,r} S_{r,k}}{S_{j,k}}+ \frac{S_{i,r} S_{r,k}}{S_{k,i}}] \
\earray
 but {\em it vanishes on using the four index formula} Eq(\ref{ssss}), whereby $ \tilde{\mu}(i,j,k) $ is antisymmetric in all three indices. 

One more beautiful property of $\mu$  is:  
\barray
\mu(i;j,k)-\mu(i;j,l) = Y_{i,l}- Y_{i,k} + \frac{S_{i,j} S_{i,k}}{S_{k,j}}+ \frac{S_{i,j}S_{i,l}}{S_{j,l}} 
= \mu(i;l,k), \label{mudiff}
\earray
where the last line follows  again from the use of the four index formula Eq(\ref{ssss}). Thus the total number of independent constraints of the $\mu$ type are only (d-2) in number\cite{overdetermined}, instead of the apparently huge number $^dC_3$. We may choose them most simply as

\beq
\mu(1;2,3) =0,  \;
...\mu(1;2,d-1) =0,  \;
\mu(1;2,d) =0.   \label{muset}
\eeq
The reader can verify that all other constraints of the type $\mu(i,j,k)=0$ are obtainable from these $d-2$ by using Eq(\ref{mudiff}) and the antisymmetry of $\tilde{\mu}(i,j,k)$.                                                                              

{\bf Constraints, variables  and consistency:} We now recount  the number of available variables versus the constraints and show how 
generic matrices of Type I can be constructed. Firstly we  construct the $S$ matrix satisfying the $(d-2)(d-3)/2$ constraints Eq(\ref{setoffives}): this can be done by
assigning arbitrary values to  $(2 d-3)$ matrix elements of $S$ and computing the rest from the constraints. A particularly convenient choice is to assign  values to $S_{1,j}$ and $ S_{2,j}$ for  $\{2,3\}\leq j\leq d$, whereby the constraints Eq(\ref{setoffives}) reduce to {\em linear equations} for the other matrix elements.  

Having determined the $S$ matrix, we next construct  the $\alpha$ matrices. A straightforward strategy is to assign arbitrary values to the  $d+2$ variables $\{a_j\}, A_1, A_2$, and  then   using the $d-2$ constraints Eq(\ref{muset}) as linear equations for the remaining $A_j$. Thus the total number of freely assignable variables for constructing the $\alpha$ matrix is $3 d- 1$.

 We can next determine a  $\beta$ matrix  by  assigning arbitrary values to the set of $d+1$ variables  $\{b_j\}$ and one of the $B_j$'s (say $B_1$),  and using  the linear Equations(\ref{xidetail}) for  $\xi_{i,1}$ to determine the rest. Thus $d+1$ is the number of independent $\beta$ type matrices of Type I for a given $\alpha$.

By construction the resulting $\beta$ and $\alpha$ matrices commute for all $x$. It is further clear  that the Equations(\ref{yeqy}) are satisfied identically. Hence it follows that $\mu(i;j,k)=0$ for all triples is satisfied whether we use $Y_{i,j}(\alpha)$ or $Y_{i,j}(\beta)$. This guarantees that starting with a matrix $\alpha(x)$ of Type I, the resulting matrix $\beta(x)$
is {\em automatically } of Type I.

We  noticed that in addition to the solutions presented above,
there are some beautiful special  solutions of these constraints Equation(\ref{rrrr}) belonging to the class of Toeplitz matrices.
From inspection and using various addition theorems, it is clear that the following class of
$R's$ satisfy this constraint identically:
\barray
R_{i,j}  = (\kappa(i) - \kappa(j))\;\;  \mbox{ or} \;\; R_{i,j} =\sin(\kappa(i) - \kappa(j)),
\earray
where $\kappa(j)$ is an arbitrary function of its argument. 

We plan to return to the problem of Hermitean as well as symplectic matrices in  a future work\cite{hermitean}.  One expects 
similar results to the ones presented here, but with more elaborate constraints.  The results in $d=3$  \cite{emil} are contained in our present ones, and correspond to the simple case of requiring  $\mu(1;2,3)=0$; this result  is sufficient to make the discriminant vanish at points in $x$. For higher $d$ the analysis of the discriminant is more difficult, however
examination of several  examples of Type I matrices leads us to conjecture that these {\em always lead to level crossings}\cite{html}.  A direct  algebraic proof involves examining the condition for vanishing of the  discriminant of the matrices, and is currently being pursued.

An important issue concerns {\em translation invariance} of the results for Type I matrices.
It is not obvious that a matrix of Type I  remains so if we shift  $x$ by a constant  and absorb the change into the diagonal piece $a$ by rediagonalizing using a suitable orthogonal transformation. 
 We have verified   that the results {\em do possess} this  translation invariance in the parameter $x$, both  numerically ( for small $d$) and analytically. A detailed proof using the parameter derivatives of the constraints Eqs(\ref{muset},\ref{ssss}), using the Pechukas flow equations \cite{yukawa} is essentially complete, and will be published separately. In future work we hope to address several physical models using these constraints ( and some obvious variants), to check for their compliance.

 I acknowledge  support from the NSF  grant  DMR 0408247.



\begin{references}
\bibitem{lenz}    The Runge Lenz vector for the Hydrogen atom is  a celebrated example of a dynamical symmetry in quantum theory,  giving an explanation of the ``permanent degeneracy'' between different $|\vec{L}|$ eigenstates. It is also untypical in that it can be cast into  a group theoretic framework, leading to  $SO(4)$ dynamical symmetry group of that problem. Our concern below is more with level crossings and their relationship with dynamical symmetries, where such a group structure is not available.
\bibitem{identitymatrix} It is clear that the commutators of interest are unchanged by adding a constant times the identity matrix to $\alpha$ or $\beta$, the equations that follow reflect this invariance.
\bibitem{alphadep}   The $\alpha$ dependence of the $Y_{i,j}$ is suppressed for brevity.
\end{references}
\end{document}